\documentclass[twoside,journal]{IEEEtran}
\usepackage{amsmath,amsfonts}
\usepackage{algorithmic}
\usepackage{algorithm}
\usepackage{array}
\usepackage[caption=false,font=normalsize,labelfont=sf,textfont=sf]{subfig}
\usepackage{textcomp}
\usepackage{multirow}
\usepackage{booktabs}
\usepackage{makecell}
\usepackage{bbm}
\usepackage{stfloats}
\usepackage{url}
\usepackage{verbatim}
\usepackage{graphicx}
\usepackage{cite}
\usepackage{CJKutf8}
\usepackage{stfloats}
\begin{document}
\begin{CJK}{UTF8}{gbsn}

\title{DECAN: A Denoising Encoder via Contrastive Alignment Network for Dry Electrode EEG Emotion Recognition}

\author{Meihong Zhang, Shaokai Zhao, Shuai Wang, Zhiguo Luo, Liang Xie, Tiejun Liu, Dezhong Yao, ~\IEEEmembership{Senior Member,~IEEE,} Ye Yan and Erwei Yin
\thanks{This work was supported in part by the grants from the National Natural Science Foundation of China under Grant 62332019, 62076250 and 62406338, the National Key Research and Development Program of China (2023YFF1203900, 2023YFF1203903). (Corresponding authors: Shaokai Zhao; Erwei Yin.)}

\thanks{Meihong Zhang, Shuai Wang are with School of Life Science and Technology, University of Electronic Science and Technology of China, Chengdu, China. And Defense Innovation Institute, Academy of Military Sciences (AMS), Beijing, China. (e-mail: zmh@std.uestc.edu.cn; tjwangshuai1234@163.com).}
\thanks{Shaokai Zhao, Zhiguo Luo, Liang Xie, Ye Yan, Erwei Yin are with Defense Innovation Institute, Academy of Military Sciences (AMS), Beijing, China. (e-mail: lnkzsk@yeah.net; 963619079@qq.com; xielnudt@gmail.com; yynudt@126.com;
yinerwei1985@gmail.com).}
\thanks{Tiejun Liu, Dezhong Yao are with The Clinical Hospital of Chengdu Brain Science Institute, MOE Key Lab for Neuroinformation, University of Electronic Science and Technology of China, Chengdu, China. And School of Life Science and Technology, Center for Information in Medicine, University of Electronic Science and Technology of China, Chengdu, China. (e-mail:
liutiejun@uestc.edu.cn; dyao@uestc.edu.cn).}}

\markboth{IEEE xxx xxx xxx xxx, ~Vol.~xx, No.~x, xx~2024}%
{Zhang \MakeLowercase{\textit{et al.}}: DECAN: A Denoising Encoder via Contrastive Alignment Network for Dry Electrode EEG Emotion Recognition}


\maketitle

\begin{abstract}
EEG signal is important for brain-computer interfaces (BCI). Nevertheless, existing dry and wet electrodes are difficult to balance between high signal-to-noise ratio and portability in EEG recording, which limits the practical use of BCI. In this study, we propose a Denoising Encoder via Contrastive Alignment Network (DECAN) for dry electrode EEG, under the assumption of the EEG representation consistency between wet and dry electrodes during the same task. Specifically, DECAN employs two parameter-sharing deep neural networks to extract task-relevant representations of dry and wet electrode signals, and then integrates a representation-consistent contrastive loss to minimize the distance between representations from the same timestamp and category but different devices. To assess the feasibility of our approach, we construct an emotion dataset consisting of paired dry and wet electrode EEG signals from 16 subjects with 5 emotions, named PaDWEED. Results on PaDWEED show that DECAN achieves an average accuracy increase of 6.94$\%$ comparing to state-of-the art performance in emotion recognition of dry electrodes. Ablation studies demonstrate a decrease in inter-class aliasing along with noteworthy accuracy enhancements in the delta and beta frequency bands. Moreover, an inter-subject feature alignment can obtain an accuracy improvement of 5.99$\%$ and 5.14$\%$ in intra- and inter-dataset scenarios, respectively. Our proposed method may open up new avenues for BCI with dry electrodes. PaDWEED dataset used in this study is freely available at https://huggingface.co/datasets/peiyu999/PaDWEED.

\end{abstract}


\section{Introduction}
\IEEEPARstart{E}{motion} plays a crucial role in human behavior, decision-making, social interactions, and overall well-being. In recent years, emotion recognition has sparked significant interdisciplinary interest from fields ranging from psychology to engineering~\cite{DBLP:journals/inffus/EzzameliM23}, based on multiple physiological signals~\cite{DBLP:journals/tii/PengLKNLC23, DBLP:conf/icassp/XiaoQTH23,kipli2022gsr,DBLP:conf/mmsp/Mirmohamadsadeghi16}. The cause appears to be, in part, that the signals mentioned above are difficult to disguise, and hence they reflect genuine emotional states~\cite{DBLP:journals/taffco/DingHXLZ21}. Electroencephalogram (EEG) is considered as an effective and reliable neural signal that carry the information about different emotions, as it objectively records our brain activity, which serves as the central nervous system for emotion processing~\cite{DBLP:journals/taffco/Garcia-Martinez21}.

EEG signals are extremely weak, which are easily susceptible to environmental noise and other electrophysiological signals \cite{yin2024motion}. Wet electrodes, regarded as the gold standard for EEG devices and clinical EEG recordings~\cite{DBLP:journals/titb/HeCPSCJK23,freire2010impedance}, offer the relative higher signal-to-noise ratio, have yielded numerous noteworthy research findings in the field of brain-computer interface~\cite{DBLP:journals/taffco/PengWKNLC22,DBLP:journals/taffco/ShenLHZS23,DBLP:conf/icira/ZhangWJXWZ19}. Although they have achieved many noteworthy research results in the field of brain-computer interface, the complex preparation and cleaning procedures have severely limited the practical application of BCI in real-world scenarios, prompting exploration into dry electrode systems~\cite{DBLP:journals/neuroimage/KamGSPHHDK19,hinrichs2020comparison,lan2023investigating}. 

However, the low signal-to-noise ratio of dry EEG systems presents a challenge for portable emotion recognition applications. Inspired by the achievements by wet electrode EEG systems, we propose to enhance the emotion recognition accuracy of dry EEG recordings by leveraging insights gained from wet EEG data. Our novel approach involves training a Denoising Encoder using a Contrastive Alignment Network (DECAN) to extract task-specific information embedded within dry electrode signals with the assistance of wet electrodes which is not necessary in test phase. The main contributions of this study can be summarized as follows: 
\begin{itemize}
\item We propose a Denoising Encoder by Contrastive Alignment Network (DECAN) to enhance the recognition performance of dry electrode signals by leveraging the knowledge learned from wet electrode data, which is comprised of two partially-shared deep neural networks (DNN) for efficient emotional feature extraction and a feature alignment contrastive learning strategy. 
\item We construct a new dataset consists of Paired Dry and Wet Electrode EEG Data (PaDWEED) collected from 16 subjects under identical video stimuli and experimental conditions to validate the effectiveness of our DECAN in enhancing dry-electrode performance. Additionally, we provide baseline performance metrics for PaDWEED which can also be set as a benchmark for the dry electrode EEG emotion recognition task.
\item Experiments on PaDWEED show that our proposed DECAN achieves state-of-the-art results on dry electrode EEG emotion recognition, and inter-subject improvements observed in both the intra- and inter-dataset feature alignment tasks demonstrate its ability to overcome more challenging scenarios.
\end{itemize} 

The layout  of the paper is as follows. Section 2 provides a review of previous research on emotion recognition using dry electrodes. Additionally, a concise overview of existing EEG emotion recognition databases is also presented. In Section 3, a detailed explanation of the materials and protocols used for our dataset constructing is provided. Section 4 and Section 5 introduce our proposed DECAN model and extensive experiment results. Section 6 and 7 are dedicated to the discussion of the findings and the conclusion of the study respectively.


\section{Related Works}
\subsection{EEG emotion recognition based on dry electrode}
The portability of dry electrode EEG system enables the potential for brain-computer emotion recognition applications in daily life. However, its low signal-to-noise ratio results in poor system performance. Currently, two mainstream solutions have been proposed to mitigate this issue. From a hardware perspective, enhancements are made to the dry electrode EEG acquisition system. This includes optimizing the electrode structure and replacing the electrode material to reduce contact impedance, thereby enhancing the signal-to-noise ratio of the acquired signal. So far, there are several mature commercial dry electrode devices available for EEG data acquisition in emotion recognition, such as the DSI-24, OpenBCI, EMOTIV EPOC and so on. In another study, a four-channel textile cap was designed with dry electrodes secured by an ultra-soft gel holder while introducing stylish and ergonomic design features to enhance wearability and comfort. The average binary emotion classification accuracy was found to be 81.32$\%$ among five healthy elderly participants~\cite{fangmeng2020textile}. 

From a software perspective, the performance of the dry electrode EEG system has been enhanced by optimizing the recognition algorithm, including feature extraction, classifier design. Lakhan et al. recruited 43 subjects to watch pre-labeled emotional visual stimuli, while using OpenBCI and Empatica4 to collect EEG and peripheral physiological signals. Through a classification algorithm based on K-means clustering, they achieved accuracy rates of 70$\%$ for arousal and 67$\%$ for valence~\cite{lakhan2019consumer}. Katsigiannis et al. simultaneously collected EEG and Electrocardiogram (ECG) data from subjects while they were viewing movie clips. When using unimodal features, the SVM-RBF classification method achieved higher accuracy rates for valence, arousal, and dominance using EEG features compared to ECG features. The accuracy rates for valence, arousal, and dominance were 62.49$\%$, 62.17$\%$, and 61.84$\%$ respectively~\cite{DBLP:journals/titb/KatsigiannisR18}. Javaid et al. employed the SVM-RBF algorithm to classify EEG data collected using OpenBCI, obtained an accuracy of 87.62$\%$ for arousal and 83.28$\%$ for valence~\cite{DBLP:conf/iconip/JavaidYSASK15}. Lan et al. collected EEG signals from individuals with major depressive disorder and healthy control subjects. They utilized the topological information among EEG channels for emotion recognition and depression detection. The study evaluated the promising ability of the emotional EEG patterns in distinguishing individuals with depression from the healthy control group~\cite{lan2023investigating}. Xu et al. put forward a novel framework for emotion recognition in VR emotional scenes using EEG signals. They employed feature extraction techniques in the time domain, frequency domain, and spatial domain from the EEG data. The extracted features were then utilized to train an ensemble model using the Model Stacking approach, combining gradient boosting decision tree, random forest, and SVM models. The resulting average accuracy achieved for classifying positive and negative emotions was approximately 81.30$\%$~\cite{xu2019emotion}. However, given that research on dry electrode emotion recognition is still in its nascent stages, the methods adopted in the aforementioned studies primarily rely on traditional signal processing and machine learning algorithms. The extracted EEG features are relatively shallow, and primarily focusing on the recognition of emotion dimension models. In contrast, this study focuses on dry electrode EEG discrete emotion recognition, employing deep learning methods to leverage wet electrode signals with relatively higher signal-to-noise ratio to enhance the performance of dry electrode EEG emotion recognition.

 \begin{table*}[!t]
\caption{Publicly available databases for EEG-based emotion recognition}
\label{tab1}
\setlength{\tabcolsep}{3pt}
\renewcommand{\arraystretch}{1}
\begin{tabular}{cccccc}
   \toprule
    Database & Stimuli & Participants & Emotion states & EEG Electrode & Motivation \\
   \midrule
     \thead{MAHNOB-HCI} &  \thead{20 video clips} & \thead{27} & \thead{Valence, Arousal, Dominance, \\Predictability, Anger, Anxiety, \\Fear, Sadness, Disgust, Neutrality, \\Surprise, Amusement, Joy} & \thead{Wet} & \thead{Multimodal synchronization database.} \\
     
     \thead{DEAP} &  \thead{40 music video clips} & \thead{32} & \thead{Valence, Arousal, Dominance, \\Liking, Familiarity}  & \thead{Wet} & \thead{Using music video as stimulus\\ for the first time.} \\
     \thead{ASCERTAIN} & \thead{36 movie clips} & \thead{15} & \thead{Negativity, Neutrality, Positivity} & \thead{Wet} & \thead{Exploring the influence of individual\\ traits on emotion classification using\\ multimodal physiological signals.} \\
     
    \thead{MPED} & \thead{28 video clips} & \thead{23} & \thead{Anger, Fear, Sadness, Disgust, \\ Neutrality, Funny, Joy} & \thead{Wet} & \thead{Eliminating the impact of cultural\\ background differences on multimodal \\databases through large-scale\\ rigorous screening.}\\
    
   \thead{SEED}& \thead{15 film clips} & \thead{15} & \thead{Negativity, Neutrality, Positivity} & \thead{Wet} & \thead{Focusing on the stability \\of multimodal emotional\\ physiological signals.}\\
   
   \thead{CASE}& \thead{8 film clips} & \thead{30} & \thead{Valence, Arousal; \\Amusing, Boring, Relaxing and Scary} & \thead{Wet} & \thead{Real-time annotation.}\\
   

    \thead{DREAMER}& \thead{18 video clips} & \thead{23} & \thead{Valence, Arousal, Dominance} & \thead{Dry} & \thead{First database containing EEG \\and ECG signal recordings\\ from low-cost, off-the-shelf, \\portable wireless devices. }  
   \\
   
    \thead{\textbf{PaDWEED (Ours)}}& \thead{\textbf{25 film clips}} & \thead{\textbf{16}} & \textbf{\thead{Anger, Fear, Sadness, \\Happiness, Neutrality}} & \thead{\textbf{Dry, Wet}}& \textbf{\thead{Multimodal database for \\discrete emotions, using both\\ dry and wet electrodes \\to collect EEG signals \\ for the first time.}}\\   
   \bottomrule
\end{tabular}
\label{tab1}
\end{table*}

\subsection{Available databases for EEG-based emotion recognition}
There has been a consecutive release of  emotion databases that containing EEG, within its range this decade, which each of them serves to distinctive experimental motivation~\cite{DBLP:journals/cbm/RahmanSHHIHQM21}. Specific information of these databases has been reported in Table~\ref{tab1}. As an early released multimodal public dataset containing both physiological responses and facial expressions, MAHNOB-HCI has garnered significant attention~\cite{DBLP:journals/taffco/SoleymaniLPP12}，and numerous studies have been conducted to validate algorithm performance using this dataset~\cite{DBLP:conf/icassp/RayatdoostR020,2022Emotion,DBLP:journals/bspc/ZhangCWX22}. The DEAP dataset, released during the same period, explores the possibility of classifying emotions induced by music videos, where this type of stimulus has never been explored before~\cite{DBLP:journals/taffco/KoelstraMSLYEPNP12}. Also from the perspective of selecting stimulus materials, Song et al. eliminated the impact of cultural dependence on the induction of discrete emotions through large-scale and rigorous screening of materials. Finally, they elaborately selected 28 videos as standardized elicitation samples and recorded the physiological responses of subjects while watching the above videos. signal, thereby constructing the MPED database~\cite{DBLP:journals/access/SongZLZZC19}.

Changes in participants' physiological states over time inevitably lead to a decline in emotion classification performance, posing challenges for practical applications. To address this issue, Zheng et al. introduced a novel dataset called SEED and aimed to discover stable patterns within repeated sessions of the same participant. The results obtained indicated that neural patterns during training and across different training instances were relatively stable~\cite{DBLP:journals/tamd/ZhengL15}. Taking a similar perspective on the influence of time on emotions, Sharma et al. proposed and developed the CASE database, which comprises continuous annotations of emotions and data from multiple physiological sensors~\cite{DBLP:journals/corr/abs-1812-02782}. They posited that emotions are dynamic phenomena that evolve over time in response to stimuli~\cite{DBLP:journals/taffco/SoleymaniAFP16}, thus emphasizing the examination of the temporal nature of emotional changes.

Taking into consideration that emotions are highly subjective phenomena and are influenced by various factors including personality, background, and psychological factors, Subramanian et al. proposed the ASCERTAIN database, which is the first database to link personality traits and emotional states through physiological responses. The database includes big-five personality scales, self-ratings of emotions from 58 users, as well as their physiological and facial activity  data~\cite{DBLP:journals/taffco/SubramanianWAVW18}. In addition, considering the integration of affective computing with various daily applications, the DREAMER dataset has been proposed, which is a database consisting of EEG and ECG signal records collected by portable devices, aiming at identifying the affective state after each stimuli, in terms of valence, arousal, and dominance~\cite{DBLP:journals/titb/KatsigiannisR18}. 


In this paper, to better support our research, we develop a new EEG dataset which includes Paired Dry and Wet electrode EEG Data (PaDWEED). This dataset is obtained from 16 participants who watch the same film clip stimuli in two separate experiments. In addition to EEG, physiological measurements such as ECG, electrooculogram (EOG), blood volume pulse (BVP), galvanic skin response (GSR), respiration (RSP), and skin temperature (SKT) are also recorded. To the best of our knowledge, this dataset represents a pioneering effort by offering both wet and dry electrode EEG signals collected from the same set of participants, which is valuable for understanding the underlying mechanisms and developing classification algorithms for emotion recognition using dry electrodes.

\section{DATASET CONSTRUCTION}

We build a new EEG emotion dataset with Paired Dry and Wet Electrode EEG Data (PaDWEED) which differs from existing publicly available datasets, to support our research. In our experiment, we recruit the same subjects to participant two separate sessions of both dry and wet electrode system experiments. For each subject, the order of the experiments is not predetermined and there is a minimum interval of two weeks between the two sessions, or a longer duration. A summary of the compiled data is provided in Table~\ref{tab2}.

\subsection{Ethics statement}
All subjects participated in the experiment on the basis of understanding the experimental procedures and equipment safety, and signed an informed consent form before the experiment. The experiment is conducted in accordance with the guidelines of the Declaration of Helsinki and is approved by the ethics committee of the University of Electronic Science and Technology of China.

\begin{figure}[!t]
\centerline{\includegraphics[width=0.8\columnwidth]{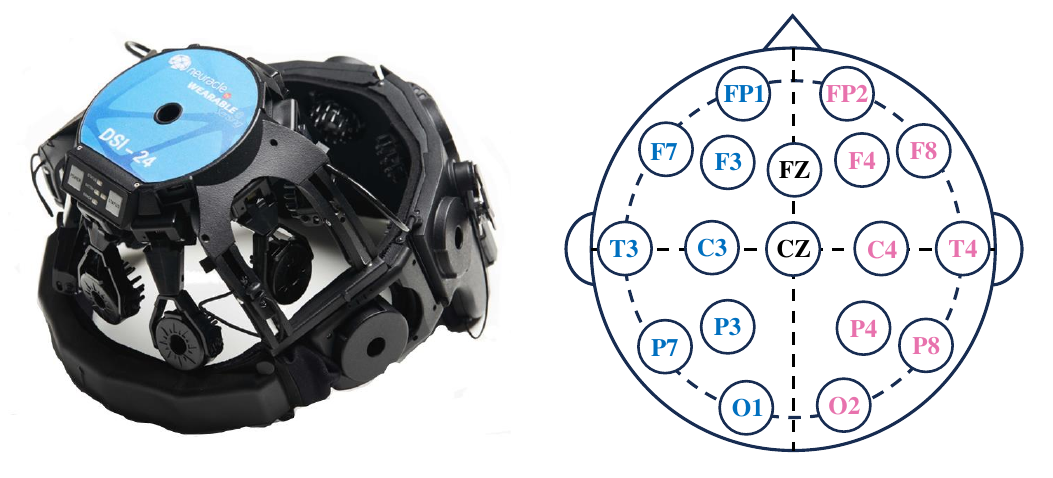}}
\caption{The DSI-24 EEG cap and the sensor layout with 18 channels.}
\label{Fig.1}
\end{figure}

\subsection{Experiment Setup}

\begin{figure}[!b]
\centerline{\includegraphics[width=\columnwidth]{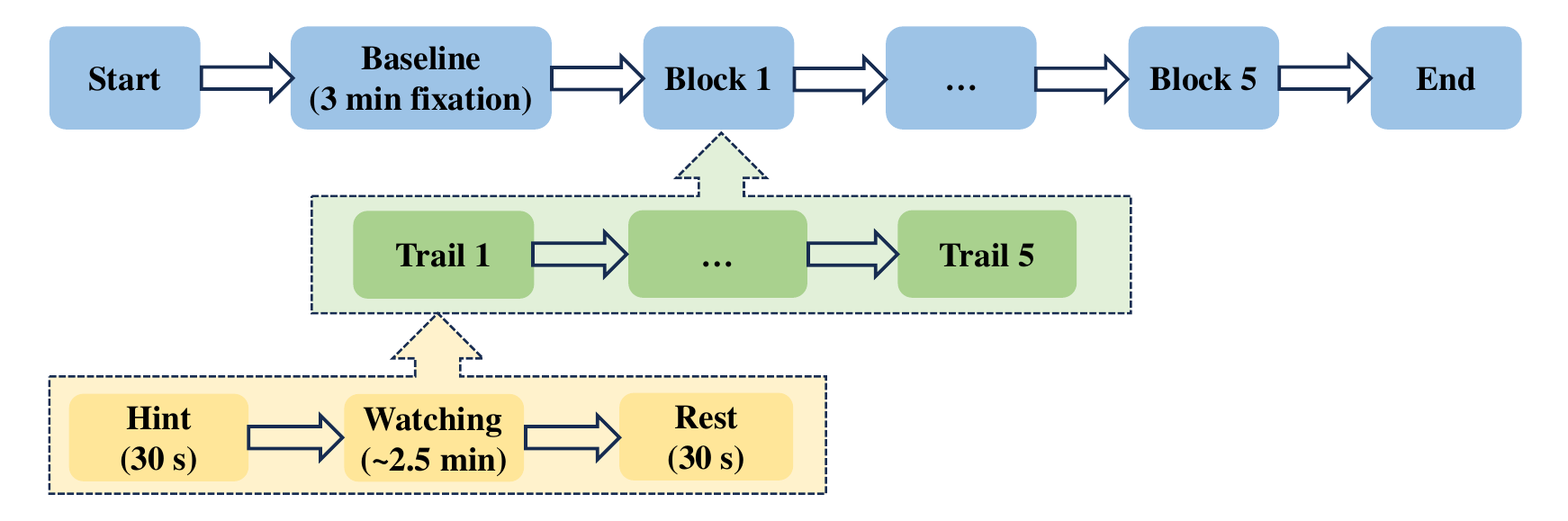}}
\caption{The protocol used in both wet and dry electrode emotion experiments.}
\label{Fig.2}
\end{figure}

\textbf{Stimulus selection}. In our emotion experiments, we select predefined Chinese movie clips as stimuli for emotion elicitation~\cite{wang2023increasing}. This choice is made considering the potential influence of native cultural factors on emotion elicitation in experimental settings~\cite{DBLP:conf/embc/ZhengDL14,DBLP:conf/ner/WuSZLY17}. Specifically, the 25 videos correspond to the emotions of anger, fear, sadness, happiness, neutrality, with an equal number of stimuli in each emotion category. The length of each edited clip is approximately 2.5 minutes. The selection of these video materials is based on three criteria: (a) stability of video content, (b) retention of characters in the scenes (excluding neutral materials), and (c) the absence of simultaneous presence of positive and negative emotions. Each video clip is edited to ensure consistency between the scenes and emotional content throughout the entire performance.


\textbf{Materials}. Two PCs are used, one for stimulus presentation, positioned approximately one meter in front of the user, and another for recording data, allowing the experimenter to verify the recorded sensor data. In the wet electrode system experiment, an ESI NeuroScan System2 with a 62-channel active AgCl electrode cap is used to  collect  EEG  data  with a sampling frequency of 1000 Hz. In the dry electrode experiment, we utilize the Dry Sensor Interface (DSI-24) and DSI-Streamer to record EEG signals at a sampling rate of 300 Hz. The sensors aree positioned according to the international 10-20 system, with the default setting of the Pz electrode as the reference. The DSI-24 EEG cap and sensor layout for 18 channels are illustrated as Fig.~\ref{Fig.1}. 

Although peripheral physiological signals are not the focus of this study, we still utilize the MP160 data acquisition system to collect ECG, GSR, PPG, RSP, and ST signals at a sampling rate of 1000 Hz, which have demonstrated promising performance in emotion estimation research, therefore may be useful for other studies. The stimulus presentation protocol is developed using MATLAB's Psychtoolbox. Synchronization markers are sent from the stimulus presentation PC to the physiological data recorder to mark the start and end of each stimulus. 

\textbf{Participants}. A total of 16 volunteers (10 males, 6 females) with an average age of 23.5 ± 1.71 years participate in both data collection experiments. All participants are recruited from the Tianjin (Binhai) Artificial Intelligence Civil-Military Integration Innovation Center and are university students. They self-reported normal vision or corrected-to-normal vision and normal hearing. Prior to the experiment, participants are informed about the experimental procedure and instructed to sit comfortably and attentively watch the upcoming movie clips without diverting their attention from the screen and to minimize any noticeable movements.

\textbf{Protocol.} Each participant performs the experiment in a session lasting approximately 75 minutes in which we define each viewing of a video clip as one trial. The experiment begins with a three-minute baseline recording during which participants are shown a fixed cross. The video clips are then divided into 5 blocks, with each block consisting of 5 trials. Each trial included a 30-second introductory prompt before each clip and a 30-second rest period after each clip. During both the introductory prompt and the rest period, participants are given the freedom to decide whether to wait or proceed to the next stage based on their own condition. Fig.~\ref{Fig.2} presents the detailed protocol. 

\section{Methodology}
\subsection{Problem setup.}
In dry electrode EEG emotion recognition systems, poor recognition performance can be attributed to lower signal-to-noise ratio, as weaker and noisier signals may hinder the detection of subtle emotion patterns and affect the overall recognition performance. Suppose ${X_w} = \left\{ {\left( {x_{wi}^j,y_{wi}^j} \right),\forall i \in \left[ {1,r} \right],\forall j \in \left[ {1,s} \right]} \right\}$ and ${X_d} = \left\{ {\left( {x_{di}^j,y_{di}^j} \right),\forall i \in \left[ {1,r} \right],\forall j \in \left[ {1,s} \right]} \right\}$ denote the set of input data and labels of wet and dry electrode signals, where $i$ and $j$ denote the sample and participant indices respectively. $r$ is the number of samples belonging to each participant and $s$ is the total number of participants. Based on previous observations indicating the presence of similar valuable information in both wet electrode and dry electrode systems, our objective is to enhance the accuracy of dry electrode EEG emotion recognition by leveraging the valuable information obtained from wet electrode EEG signals.

\subsection{Solution overview}
\begin{figure*}[!t]
\centerline{\includegraphics[width=2\columnwidth]{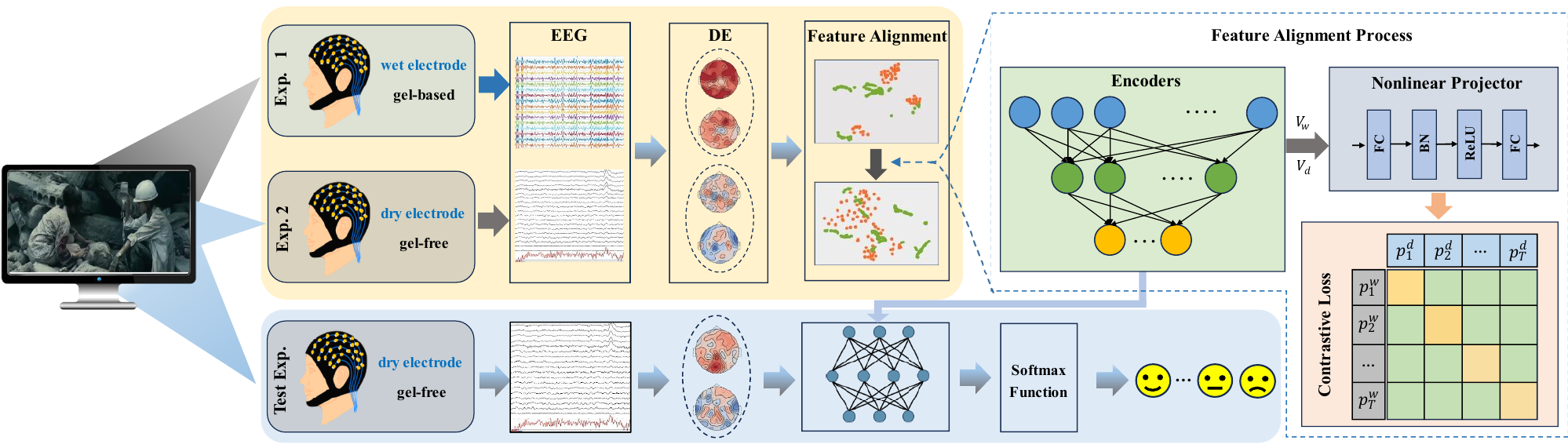}}
\caption{The architecture of our proposed framework DECAN with pairwise representation alignment with contrastive learning for emotion recognition.}
\label{Fig.3}
\end{figure*}

As illustrated in Fig.~\ref{Fig.3}, our proposed method is a contrastive learning-based architecture. Specifically, we utilize DNN models that have shown superior performance in previous analysis to learn high-level features from both wet and dry electrode EEG signals. Subsequently, we employ contrastive learning to perform pairwise representation alignment (wet versus dry).  In the following subsections, we provide a detailed description of each step in our proposed method.

\subsection{Our approach}

\textbf{Feature extraction.} Considering the impact of the length of samples on recognition performance~\cite{moctezuma2022two,DBLP:journals/taffco/ShenLHZS23}, we divide a trial into five-second non-overlapping segments to ensure a balanced number and sample length of training samples, where each segment is regarded as a model training sample. The recorded EEG signals are susceptible to contamination from noise and artifacts but not of cerebral origin. Therefore, a series of preprocessing steps are adopted in our research to improve the EEG quality. First, a bandpass filter from 1 Hz to 50 Hz is applied to the raw EEG signals acquired from both dry and wet electrodes. This filtering approach aims to preserve the frequency range of the EEG signals containing emotional information while simultaneously removing any DC offsets and high-frequency interference such as muscle activity or other artifacts. Next, a notch filter at 50 Hz is employed to eliminate the power line interference caused by AC electrical sources or poor grounding connections. Finally, the filtered signals are downsampled to 200 Hz to reduce the complexity of subsequent signal processing and accelerate the overall signal processing speed.

It is well documented that the differential entropy (DE) is of efficient EEG feature for human emotional states~\cite{DBLP:conf/embc/ShiJL13,DBLP:journals/taffco/ZhengZL19}. Hence, it is employed in this research to facilitate the subsequent classification model. Specifically, the preprocessed EEG signals are separated into five frequency band: delta (1-4 Hz), theta (4-8 Hz), alpha (8-14 Hz), beta (14-31 Hz) and gamma (31-50 Hz), then the DE for Gaussian distribution can be extracted as follows:

\begin{equation}
h\left( X \right) = \frac{1}{2}\log \left( {2{\rm{\pi }}e{\sigma ^2}} \right)
\label{eq1}\end{equation}
where $X$ denotes the Gaussian distribution $N\left( {\mu ,{\sigma ^2}} \right)$, ${\rm{\pi }}$ and $e$ are constants.

Assuming emotional states are situated in a continuous space and gradual transition, the linear dynamical system is a popular feature smoothing technique widely employed to filter out components irrelevant to emotional states~\cite{DBLP:journals/taffco/ZhengZL19,lu2015combining,zhu2014eeg}. Given the impressive performance demonstrated by this method in various studies, we also employ it to smooth the extracted DE features.

\textbf{Encoders.} Given a series of paired wet and dry electrode EEG signals ${X_w}$ and ${X_d}$ collected from same participant during the same stimulus segment, we use the extracted DE features from multiple frequency bands ${F_w} = \left\{ {f_{iw}^j,\forall i \in \left[ {1,r} \right],\forall j \in \left[ {1,s} \right]} \right\}$ and ${F_d} = \left\{ {f_{id}^j,\forall i \in \left[ {1,r} \right],\forall j \in \left[ {1,s} \right]} \right\}$ as inputs to the encoder and output high-level features ${V_w} = \left\{ {v_{iw}^j,\forall i \in \left[ {1,r} \right],\forall j \in \left[ {1,s} \right]} \right\}$, ${V_d} = \left\{ {v_{id}^j,\forall i \in \left[ {1,r} \right],\forall j \in \left[ {1,s} \right]} \right\}$. Specifically, ${F_w}$ and ${F_d}$ are extracted from the raw signal after filtering and downsampling, as described previously. Then we choose to utilize two partial weight-sharing DNN as the encoders in our approach. This particular architecture has consistently shown superior performance compared to other models in extracting paired features effectively. To promote efficient feature extraction, we have implemented weight sharing by sharing the last two linear sublayers and the projector between the two types of signals.

\begin{equation}
{V_w} = DN{N_w}\left( {{F_w}} \right)
\label{eq2}\end{equation}

\begin{equation}
{V_d} = DN{N_d}\left( {{F_d}} \right)
\label{eq3}\end{equation}

\textbf{Feature alignment.} Contrastive learning is a highly effective method for self-supervised representation learning, demonstrating significant achievements in pairwise feature learning across diverse domains \cite{DBLP:conf/icml/RadfordKHRGASAM21,DBLP:journals/natmi/DefossezCRKK23,DBLP:conf/icml/ChenK0H20}. Recent advancements in neuroscience have provided inspiration, suggesting that contrastive learning can be applied to extract subtle information from signals within the central nervous system\cite{DBLP:journals/taffco/ShenLHZS23}. This growing body of evidence highlights the potential of utilizing contrastive learning in neural research. Based on the aforementioned rationale, we have chosen to employ contrastive learning to accomplish the task of aligning features between wet and dry electrode EEG signals, details are introduced as follows.

\textbf{Projectors.} Motivated by the SimCLR framework, which suggests that nonlinear projection of encoded features can yield better performance in downstream tasks compared to linear projection or no operation \cite{DBLP:conf/icml/ChenK0H20}, we have opted to incorporate a structure comprised of fully connected layers and a ReLU layer. This configuration allows us to obtain the desired projection features, shown in Fig.\ref{Fig.3}.

\begin{equation}
{P_w} = Projector\left( {{V_w}} \right)
\label{eq2}\end{equation}

\begin{equation}
{P_d} = Projector\left( {{V_d}} \right)
\label{eq2}\end{equation}

\textbf{The contrastive loss.} Given a paired of sequences of projection features ${P_w} = p_i^w,p_{i + 1}^w,...,p_T^w$ and ${P_d} = p_i^d,p_{i + 1}^d,...,p_T^d$, where $T$ is the number of samples, $w$ and $d$ denote wet and dry electrode respectively. We define samples as positive when they originate from the same subject and different EEG collection systems, while also corresponding to the same time segment of the same emotion type. Conversely, any samples that do not meet these criteria are classified as negatives.

Contrastive learning employs a loss function that utilizes similarity as a measure to bring positive samples closer to each other while separating negative samples. Formally, the loss function for a pair of samples ${p_i^w}$, ${p_i^d}$ can be formulated as follows:

\begin{equation}
L_i^{w,d} =  - \log \frac{{\exp \left( {{{sim\left( {p_i^w,p_i^d} \right)} \mathord{\left/
 {\vphantom {{sim\left( {p_i^w,p_i^d} \right)} \tau }} \right.
 \kern-\nulldelimiterspace} \tau }} \right)}}{{\sum\nolimits_{j = 1}^T {{\mathbbm{1}_{\left[ {i \ne j} \right]}}} \exp \left( {{{sim\left( {p_i^w,p_j^d} \right)} \mathord{\left/
 {\vphantom {{sim\left( {p_i^w,p_j^d} \right)} \tau }} \right.
 \kern-\nulldelimiterspace} \tau }} \right)}}
\label{eq2}\end{equation}

\noindent
where ${\mathbbm{1}_{\left[ {i \ne j} \right]}} \in \left\{ {0,1} \right\}$, it set to 1 iff $i \ne j$. $i$ and $j$ are the sample indices in current batch. $\tau $ is the temperature scalar. And $sim\left( {p_i^w,p_j^d} \right)$ represents cosine similarity computed by

\begin{equation}
sim\left( {p_i^w,p_j^d} \right) = \frac{{p_i^w \cdot p_j^d}}{{\left\| {p_i^w} \right\|\left\| {p_j^d} \right\|}}
\label{eq3}\end{equation}

By iterating over all sample time points within the batch, we can calculate the final contrastive loss,

\begin{equation}
{L_{CL}} = \sum\limits_{i = 1}^T {L_i^{w,d}} 
\label{eq4}\end{equation}

\textbf{Total loss.} Due to the low signal-to-noise ratio of EEG as a physiological signal, we begin by conducting wet and dry electrode EEG emotion classification tasks. This crucial step ensures that the features extracted by the encoders from both wet and dry electrodes effectively contribute to emotion recognition. By doing so, we establish a foundation for feature alignment steps within the contrastive learning framework.

To consolidate the aforementioned information, we employ a joint training mechanism and update our final loss function as follows:
\begin{equation}
L = {L_W} + {L_D} + {L_{CL}}
\label{eq20}\end{equation}

\noindent
where ${L_W}$ and ${L_D}$ are the classification losses of wet electrode and dry electrode EEG signals respectively.

\section{Experiment Results}
In this section, we first present the baseline emotion recognition results derived from PaDWEED. Following this, we list the performance evaluation results of DECAN in comparison to other established solutions. Moreover, we conduct ablation experiments to analyze the influence of the contrastive learning framework on classification performance within a specific configuration and visualize the embeddings produced by our model. Lastly, we study the generalization performance of DECAN through alignment experiments across subjects and datasets (SEED V), where dry and wet electrodes come from different subjects and datasets respectively.

\subsection{Datasets}
\textbf{PaDWEED.} The PaDWEED dataset comprises two sub-datasets: (1) The dry electrode sub-dataset includes 24-channel EEG data gathered from 16 subjects. (2) The wet electrode sub-dataset encompasses 64-channel EEG data obtained from the identical group of 16 subjects exposed to the same video stimuli. This paired cross-device dataset facilitates the research of feature alignment between dry and wet electrode recordings.

\textbf{SEED V.} The SEED V dataset~\cite{liu2021comparing} comprises 62-channel wet electrode EEG recordings from 20 subjects, with each subject participating in three sessions. During each session, there are 15 trials, corresponding to 15 movie clips evenly distributed across 5 emotional states (happy, sad, disgust, fear, and neutral).

\subsection{Implementation details}
To evaluate the performance of our emotion recognition results and establish baseline classification results for the proposed database, we perform individual participant classification using well-established approaches commonly employed in prior research on affective computing. Specifically, we employee a support vector machine (SVM) classifier with a linear kernel, logistic regression (LR), and DNN. These methods have demonstrated their effectiveness in previous studies and serve as reference models in our analysis~\cite{DBLP:journals/taffco/CorreaASP21,liu2022identifying,wu2023investigation,DBLP:journals/taffco/ZhengZL19}. Specifically, regarding the LR model, we utilize the default function provided by the scikit-learn module. As for the SVM classifier, we employee the function available in the scikit-learn module with a linear kernel. To determine the optimal misclassification cost parameter C for the linear SVM, we conduct a grid search over the range of $[2^{-10}, 2^{-9}, ..., 2^{10}]$ and $[0.1, 20]$, using a step size of 0.5 for the large-step and small-step scenarios, respectively. In the DNN model utilized in this paper, we integrate three hidden layers with 128, 64, and 32 hidden units for the wet electrode system, whereas the dry electrode system utilized only the last two hidden layers. The output layer comprised five units, each corresponding to one of the five emotions mentioned above.

In our quest to optimize the model, we empirically adjust to specific hyperparameters based on preliminary results. To enhance the training process, we employ the RMSprop optimizer with the learning rate selected from $[1^{-4}, 9^{-4}]$, $[1^{-3}, 9^{-3}]$ and $[1^{-2}, 9^{-2}]$ using step size of $2^{-4}$, $2^{-3}$, $2^{-2}$ respectively for optimization. For all the experiments conducted, our model is trained for a maximum of 15,000 epochs. In addition, the sample length of 5 seconds is used to split the entire trial into segments, which facilitates consistent evaluation and benchmarking against the baseline performance of the dataset.

To strike a balance between the effectiveness of the contrastive learning component and the overall stability of the training process for the DECAN model, we make a strategic decision regarding the contrastive learning temperature scalar. Specifically, we selected a value of 0.5 as a compromise. In our efforts to enhance the quality of the projection of representations into the latent space, we conduct tuning of the number of hidden units within the projection head. Specifically, we explore a range of options, including 64, 128, 256, and 512 hidden units.

\subsection{Baseline results of PaDWEED}

\begin{figure*}[!ht]
\centerline{\includegraphics[width=2\columnwidth]{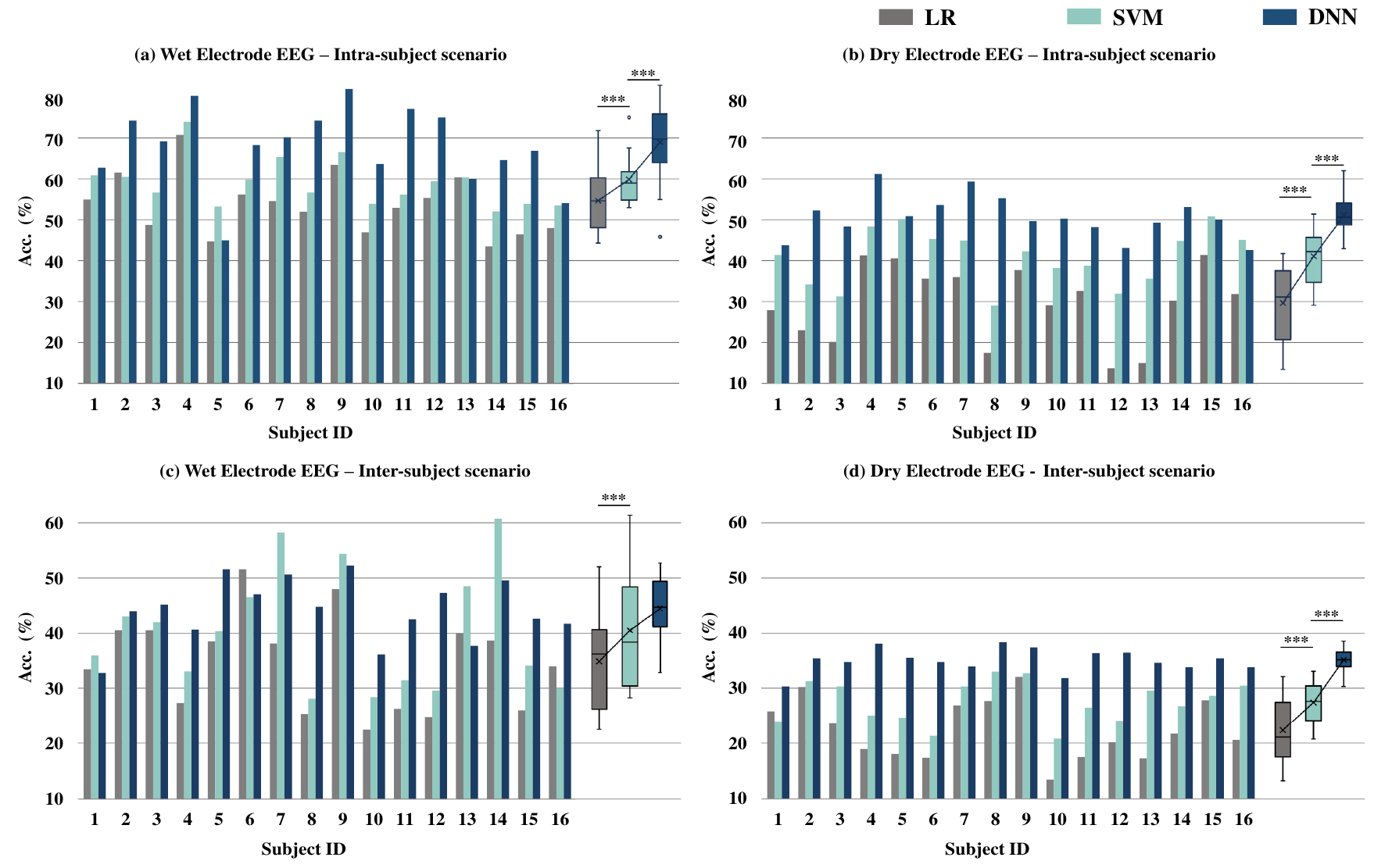}}
\caption{The baseline results of the wet-electrode and dry-electrode based emotion classification in the condition of both intra-subject and inter-subject scenarios on our proposed dataset.}
\label{Fig.4}
\end{figure*}

\textbf{Conditioned on intra-subject.} To validate the classification performance in the condition of intra-subject, we employee a leave-one-block-out (LOBO) cross-validation technique. In each step of the cross-validation process, one block of samples is held out as the test set, while the classifier is trained on the samples from the remaining blocks. This process is repeated for all blocks’ data, ensuring that each sample is used for testing exactly once. Fig.\ref{Fig.4} presents the results of three different classifiers, LR, SVM and DNN in wet and dry electrode systems. Specific quantitative results are shown in Table \ref{tab2} where the best performance in each case is highlighted in bold.

\begin{table}[!b]
\caption{Lobo test results of based on three different classifiers. The mean and standard deviation ($\%$) of accuracies are shown.}
\label{table}
\setlength{\tabcolsep}{9.5pt}
\renewcommand\arraystretch{1.5}
\begin{tabular}{ ccccccc }
   \toprule
   \multirow{2}{*}{Type} & \multicolumn{2}{c}{LR\cite{wu2023investigation}}  & \multicolumn{2}{c}{SVM\cite{DBLP:journals/taffco/CorreaASP21}} & \multicolumn{2}{c}{DNN\cite{liu2022identifying}} \\\cline{2-7} 
     & Acc  & Std  & Acc  & Std & Acc  & Std  \\\hline
   \ Wet  & 54.39 & 7.61 & 59.62 & \textbf{5.91} & \textbf{68.74} & 9.38 \\\hline
   \ Dry & 30.73 & 7.98 & 41.24 & 6.81 & \textbf{51.44} & \textbf{5.31} \\                  
   \bottomrule
\end{tabular}
\label{tab2}
\end{table}

In an effort to offer valuable insights into optimizing the recognition system for enhanced performance, a paired t-test is conducted on the recognition results of the three models, focusing on performance disparities in both the wet and dry electrode systems. The statistical test results presented in Fig \ref{Fig.4} (a) and (b) demonstrate significant differences among the models. In the wet electrode system, the DNN model exhibits superior performance compared to both the SVM and LR models, with  $p \le 0.001$. Similarly, in the dry electrode system, the DNN model achieved the best performance among the three models. However, one notable distinction from the wet electrode system is that the DNN model exhibited a more pronounced significance difference compared to the SVM model, with $p \le 0.001$. These findings highlight the effectiveness of the DNN model in both electrode systems and emphasize its superiority over alternative models.

\textbf{Conditioned on inter-subject.} We employ the Leave-One-Subject-Out (LOSO) cross-validation approach. This validation scheme involved utilizing data from 15 participants for training purposes, while the data from one participant is set aside for testing. The accuracy results for all subjects are depicted in Fig.\ref{Fig.4} (c) and (d), showcasing a comprehensive overview of the findings. A notable observation is the substantial decrease in recognition accuracy for all three models when compared to the intra-subject scenario, regardless of whether the EEG system utilizes dry or wet electrodes. Within the wet electrode EEG system, both the DNN and SVM models achieve significantly superior results compared to the LR model. However, no significant difference is observed between the DNN and SVM models. Conversely, in the dry electrode EEG system, the DNN model consistently outperforms the SVM and LR models. This difference in performance is statistically significant, with $p \le 0.001$ for both the SVM and LR models.

\begin{table}[!b]
\caption{Loso test results of based on three different classifiers. The mean and standard deviation ($\%$) of accuracies are shown.}
\label{table}
\setlength{\tabcolsep}{9.5pt}
\renewcommand\arraystretch{1.5}
\begin{tabular}{ ccccccc }
   \toprule
   \multirow{2}{*}{Type} & \multicolumn{2}{c}{LR\cite{wu2023investigation}}  & \multicolumn{2}{c}{SVM\cite{DBLP:journals/taffco/CorreaASP21}} & \multicolumn{2}{c}{DNN\cite{liu2022identifying}} \\\cline{2-7} 
     & Acc  & Std  & Acc  & Std & Acc  & Std  \\\hline
   \ Wet  & 34.87 & 8.71 & 40.43 & 10.82 & \textbf{44.31} & \textbf{5.60} \\\hline
   \ Dry & 22.53 & 5.41 & 27.52 & 3.87 & \textbf{35.11} & \textbf{2.11} \\                  
   \bottomrule
\end{tabular}
\label{tab3}
\end{table}

\textbf{Baseline results comparison.} Based on our analysis, we found a statistically significant difference in recognition accuracy between the wet electrode system and the dry electrode system in both intra-subject and inter-subject scenarios ($p \le 0.001$). Average accuracy that is 14.53$\%$ higher in intra-subject and 9.20$\%$ higher in inter-subject compared to the dry electrode system. The superior performance of the wet electrode system can be ascribed to its ability to mitigate signal noise more effectively, thereby providing more valuable information compared to the dry electrode system. Additionally, in inter-subject experiment, the wet electrode system demonstrates a decrease in emotion recognition accuracy by 16.33$\%$, while the dry electrode system exhibits an 24.43$\%$ decrease compared to intra-subject which  may  be  attributed  to the increased  variability when combining data from all subjects and differences in distribution between source and target domains.

\subsection{Performance Evaluation of DECAN}


\textbf{Main results.} We consider 5 baseline methods, all originally designed for wet-electrode EEG-based emotion recognition rather than dry-electrode systems, so we adjust the settings for these methods and ensured comparability of the results. According to Table~\ref{tab4}, DECAN consistently demonstrates superior performance across all metrics. Specifically, CDRC achieves the highest performance with an accuracy of 55.01$\%$ and an F1 score of 48.59$\%$, showing a margin of 3.57$\%$ and 4.52$\%$ over the second-best baseline method DNN. These enhancements suggest that DECAN can effectively incorporate wet electrode information into dry electrode EEG-based emotion recognition systems and hold great promise for enhancing the accuracy and robustness of such systems.

\begin{figure}[!t]
\centerline{\includegraphics[width=\columnwidth]{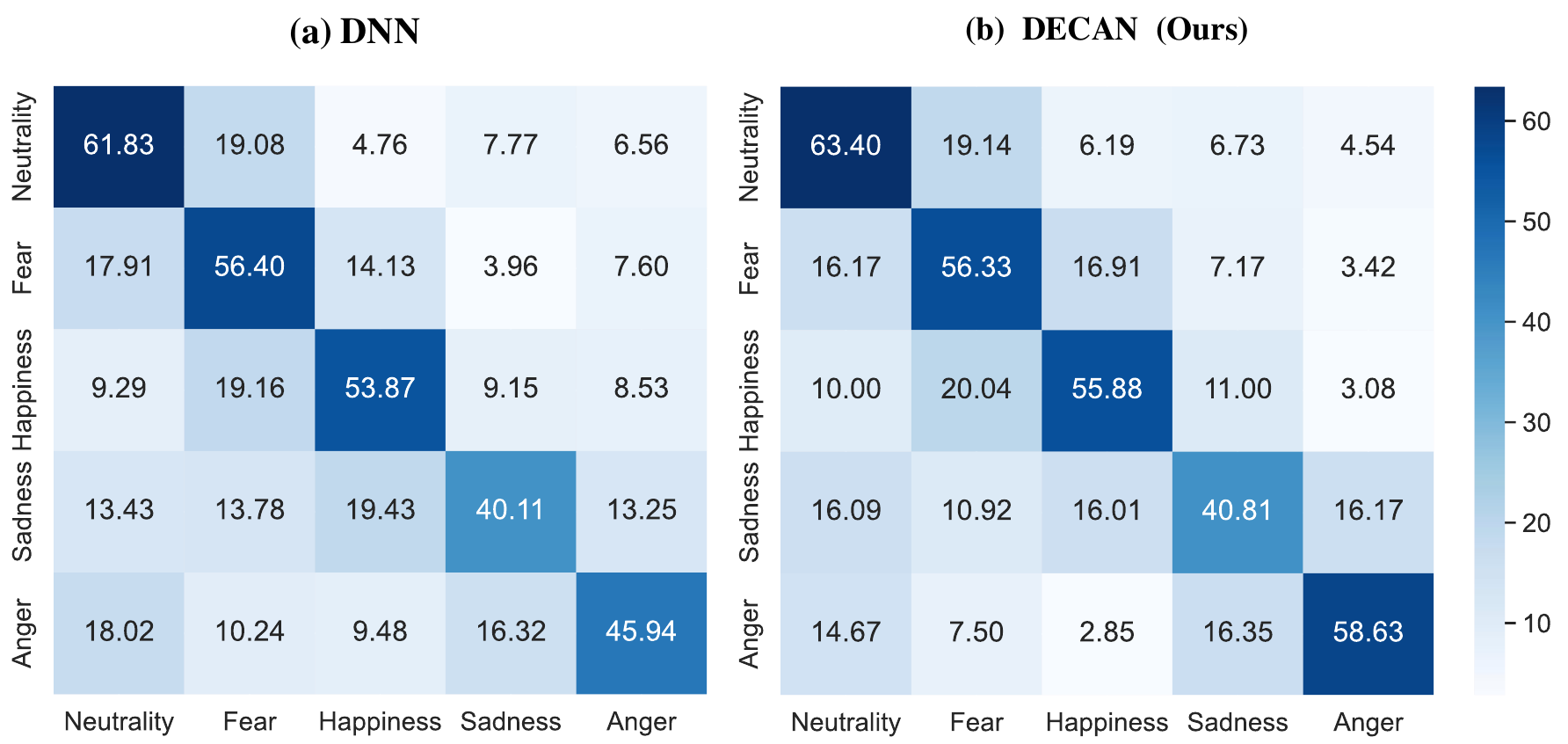}}
\caption{Confusion matrices of DNN (the second-best method) and the DECAN model for dry electrode EEG-based emotion recognition on PaDWEED dataset.}
\label{Fig.5}
\end{figure}

Fig.\ref{Fig.5} compares the classification confusion matrices of the DNN (suboptimal method) with the proposed DECAN model in the intra-subject experiment. The experiment results indicate that our model achieves higher classification accuracy for almost all emotion categories, particularly excelling in the recognition of anger among negative emotions.

\begin{figure}[!t]
\centerline{\includegraphics[width=\columnwidth]{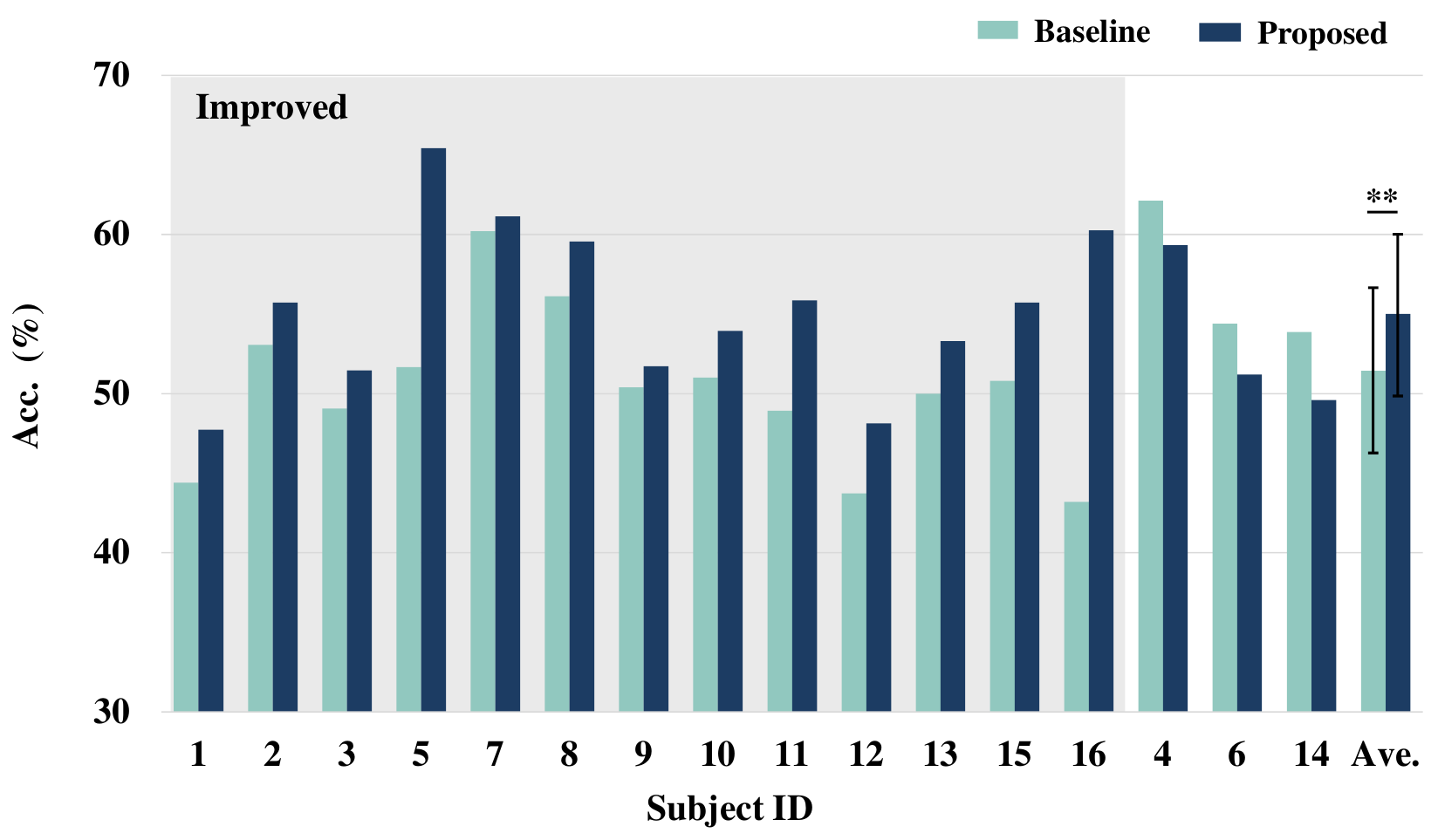}}
\caption{Emotion recognition performance of our DECAN via ablation study on our proposed dataset in the intra-subject scheme.}
\label{Fig.6}
\end{figure}

\textbf{Ablation studies.} We set the framework without the primary component contrastive learning as the baseline method in this experiment. The experimental results of the intra-subject scheme are depicted in Fig.\ref{Fig.6}. Upon analysis, it is evident that the emotion recognition accuracy of the dry electrode EEG system exhibits improvement for 13 out of the 16 subjects, with subject 5 and 16 showing enhancements of up to 13.75$\%$ and 17.07$\%$ respectively. We have further applied the paired t-test to find whether there are significant differences between the results of the baseline approach and our model at the significant level as 0.05. In our analysis, the calculated p-value was determined to be 0.01. This result suggests that there is indeed a significant difference between the results of the baseline approach and our model, demonstrating the effectiveness of the contrastive learning module. 

\begin{figure}[!b]
\centerline{\includegraphics[width=\columnwidth]{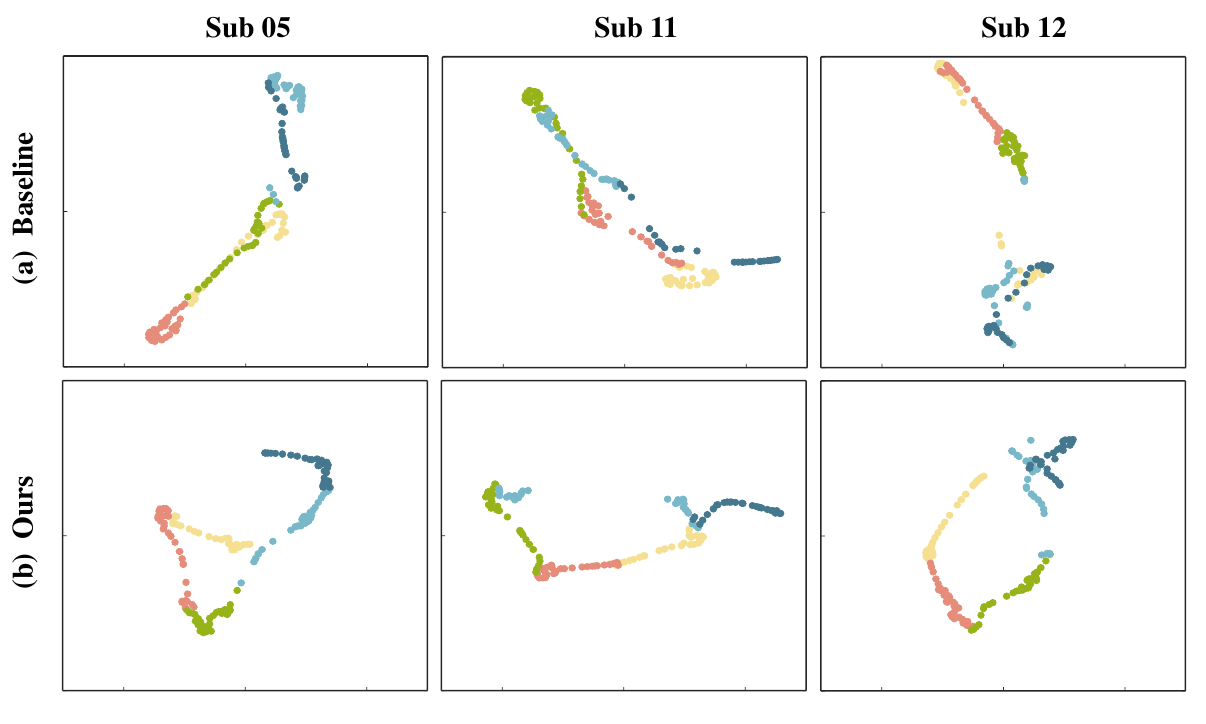}}
\caption{Visualization of latent features using t-SNE on the PaDWEED dataset. We presented the features extracted by two models: (a) Baseline, (b) DECAN. The different colors represent different emotions.}
\label{Fig.7}
\end{figure}

We further investigate the effectiveness of introducing the contrastive learning module in our proposed DECAN by visualizing the latent EEG patterns. The feature distributions of three subjects from the PaDWEED dataset are presented in a two-dimensional space using t-SNE. For each subject, two latent features with a length of 5 seconds are randomly visualized, corresponding to the two methods from a trial in the testing set. The results, as depicted in Fig.\ref{Fig.7}, clearly demonstrate that our approach, which combines the base emotion model with the contrastive learning strategy, yields more effective dry electrode EEG features.

\begin{figure}[!t]
\centerline{\includegraphics[width=\columnwidth]{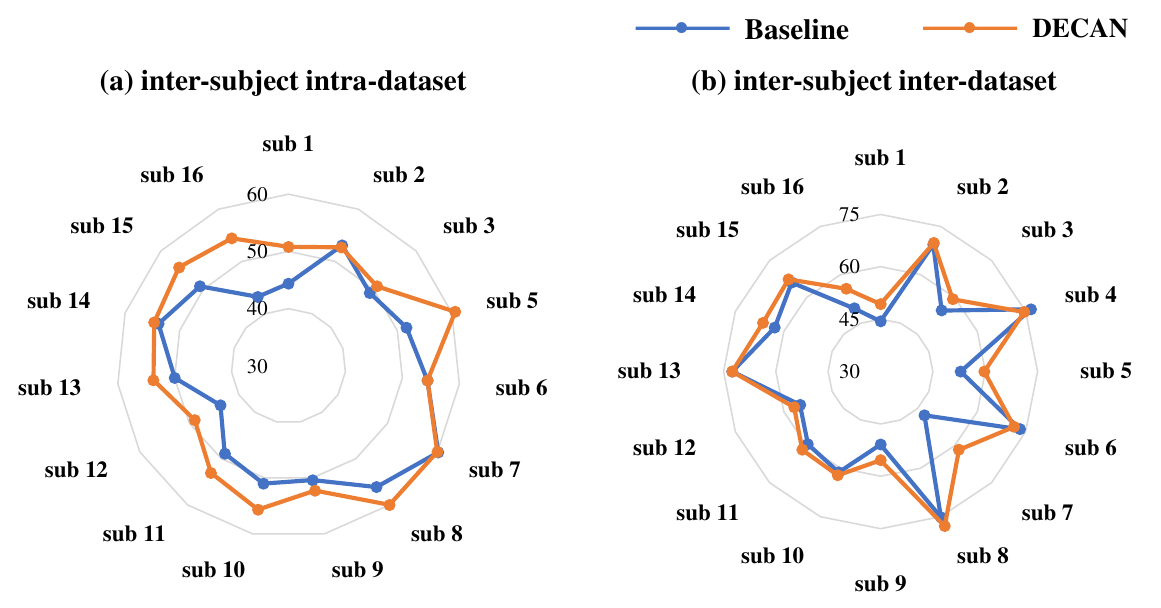}}
\caption{Generation tests of DECAN. (a) The results of dry electrode EEG emotion recognition performance of the remaining subjects assisted by wet electrode EEG signals from subject 4. (b) The results of dry electrode EEG emotion recognition performance of the subjects in PaDWEED dataset assisted by wet electrode EEG signals from SEED V dataset (subject 1).}
\label{Fig.8}
\end{figure}

\textbf{Generalization test on inter-subject intra-dataset scenario.} In this experiment conducted on the PaDWEED dataset, our objective is to assess the performance of DECAN in dry electrode emotion recognition when the wet electrode signals paired with the dry electrode ones originated from different subjects, introducing inter-subject variations. Specifically, we utilize "one-to-one" strategy, where the wet electrode EEG data from one subject are sequentially paired with the dry electrode EEG data from other subjects to construct the training set. Following this, the remaining dry electrode data trials from the subjects who provide dry electrode data in the training set are employed as the test set in sequence. For instance, we utilize the wet electrode EEG data from subject 4 as the training data, and the corresponding experiment results are illustrated in Fig.\ref{Fig.8} (a). It is evident that leveraging the same wet electrode EEG data has enhanced the accuracy for 13 out of 15 subjects, with the average accuracy showing an relative improvement of 5.99$\%$，indicating DECAN can effectively mitigate the individual differences among the subjects.

\begin{table*}[!t]
\caption{Comparison with representative methods that are widely used in the EEG analysis field. The mean and standard deviation ($\%$) of evaluation metrics are shown.}
\label{table}
\setlength{\tabcolsep}{6pt}
\renewcommand{\arraystretch}{1.4}
\begin{tabular}{p{1cm}<{\centering} p{2cm}<{\centering} p{2cm}<{\centering} p{2cm}<{\centering} p{2cm}<{\centering} p{2cm}<{\centering} p{2cm}<{\centering} p{2cm}<{\centering}}
   \toprule
    Ref. & Model& Accuracy & Precision & Recall & F1 & AUROC & AUPRC
\\
   \midrule
   \cite{wu2023investigation} & LR$^*$ & 30.73±7.98 & 27.17±8.89 & 31.03±7.62 & 26.24±7.10 & $-$ & $-$\\
   ~\cite{DBLP:journals/taffco/CorreaASP21} & SVM$^*$ & 41.24±6.81 & 35.90±7.66 & 41.16±6.86 & 34.54±6.49 & $-$ & $-$\\
   ~\cite{song2018eeg} & DGCNN$^*$ & 43.37±8.49 & 34.44±11.69 & 42.45±10.01 & 34.16±10.45 & 64.73±11.55 & \underline{44.43±9.04} \\
   ~\cite{song2022eeg} & Conformer$^*$ & 39.41±4.42 & 39.11±6.18 & 39.36±4.45 & 35.24±5.42 & 64.06±6.02 & 39.32±5.99\\
   ~\cite{liu2022identifying} & DNN$^*$ & \underline{51.44±5.31} & \underline{46.95±8.24} & \underline{51.29±5.24} & \underline{44.07±6.34} & \underline{67.43±5.28} & 43.59±5.54 \\
\hline
     & \ DECAN (Ours) & \textbf{55.01±5.07} & \textbf{51.64±8.46} & \textbf{54.91±5.01} & \textbf{48.59±6.27} & \textbf{69.45±5.73} & \textbf{45.2±6.07} \\
   \bottomrule
\multicolumn{8}{p{500pt}}{$^*$ indicates the results are obtained by our own implementation. Best results are in bold, while the second-best results are underlined.}
\end{tabular}
\label{tab4}
\vspace{-1.5em}
\end{table*}

\textbf{Generalization test on inter-subject inter-dataset scenario.} In this experiment, we additionally take into account the variations between datasets in addition to the individual variances among participants. To achieve inter-dataset pairing, the wet electrode EEG signals in the training data are sourced from the SEED V dataset (subject 1), while the dry electrode signals are obtained from the PaDWEED dataset. We specifically focused on cases with identical emotion categories (such as happy, neutral, sad, and fearful emotions) in both datasets, utilizing preprocessed 4-second EEG signals for consistency. Corresponding results are shown in Fig.~\ref{Fig.8} (b), it demonstrates that DECAN can enhance the accuracy of dry electrode EEG emotion recognition for the majority of subjects, with an average relative improvement of 5.14$\%$. This enhancement indicates that DECAN can address the challenges posed by cross-dataset scenarios and effectively boost the performance of dry electrode emotion recognition utilizing non-homologous wet electrode datasets.

\section{Discussion}
Historically, there have been two mainstreams regarding improving the emotion recognition performance of dry electrode systems, refining prototyping workflows~\cite{lan2023investigating,lakhan2019consumer,DBLP:conf/iconip/JavaidYSASK15,xu2019emotion} and optimizing hardware~\cite{fangmeng2020textile}. Here, to address the limitation of challenging hardware advancements in the short term, we focused on the former one and proposed a contrastive learning-based method to enhance the dry electrode EEG data emotion recognition performance. Unlike conventional methods that primarily focused on designing encoders to extract efficient EEG features from dry electrode systems \cite{lan2023investigating,lakhan2019consumer,DBLP:conf/iconip/JavaidYSASK15}, which often heavily depended on the quality of dry electrode EEG signals, our approach takes into consideration the challenging aspect of extracting efficient features from dry electrode EEG signals with relatively low signal-to-noise ratio. To overcome this challenge, we proposed to leverage the advantages offered by wet electrode EEG systems to enhance the emotion recognition capabilities of dry electrode systems. 


\begin{figure}[!b]
\centerline{\includegraphics[width=0.95\columnwidth]{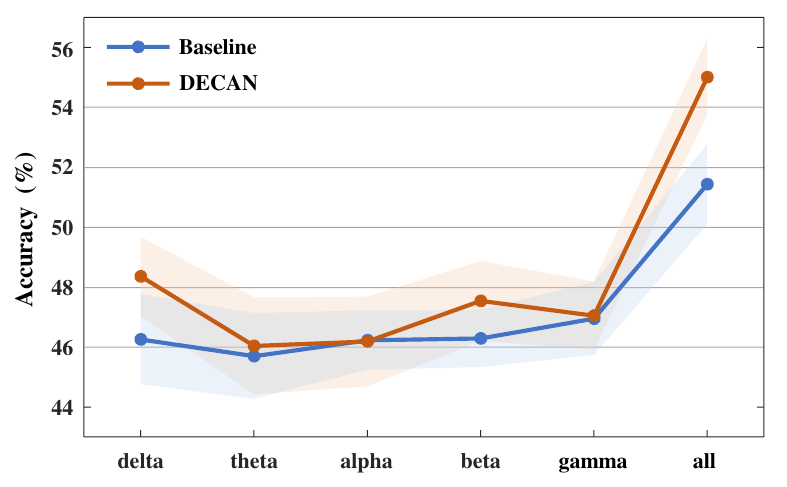}}
\caption{Emotion recognition performance of our DECAN via ablation study on our proposed dataset in the intra-subject scheme.}
\label{Fig.9}
\end{figure}

The intra-subject analysis on the PaDWEED dataset demonstrates that our proposed DECAN effectively improves the emotion recognition performance of dry electrode EEG systems by leveraging knowledge obtained from wet electrode EEG systems. This finding supports the existence of shared features between dry and wet electrodes when performing the same task, which is consistent with previous research ~\cite{DBLP:journals/neuroimage/KamGSPHHDK19,hinrichs2020comparison,lan2023investigating}. It also highlights the potential of contrastive learning methods in uncovering this specific information. Moreover, the inter-subject feature alignment analysis reveal promising results in improving the emotion recognition performance of an individual's dry electrode EEG by utilizing wet electrode EEG data from other subjects and even from non-homologous dataset. This suggests that DECAN proposed in this study can effectively align EEG signals across devices and subjects. Consequently, it eliminates the need for each subject to undergo the corresponding wet electrode experiment, significantly reducing the complexity of the overall experimental setup. 

The contrastive learning architecture in DECAN has been shown to be effective in improving dry electrode EEG emotion recognition performance. Here, we further investigate its impact on the accuracy obtained with different frequency sub-band signals for each of the 16 participants and visualize the average results in Fig.~\ref{Fig.9}. In the baseline results, the recognition accuracy is notably higher in the high-frequency bands, especially the gamma band, which is consistency with previous studies~\cite{9760385, 9597546, 9321519}. Upon employing the DECAN model with the contrastive learning module, enhancements are observed in the recognition accuracy of the full-band signal. Moreover, there has been a significant improvement in the recognition accuracy of the delta and beta bands, while this outcome of other sub-bands has remained stable. This suggests that delta and beta bands also play a substantial role in dry electrode EEG emotion recognition, aligning with findings from prior research~\cite{zhang2023eeg}.

Lastly, to our knowledge, PaDWEED is the first database which contains paired dry and wet electrode EEG data from collected with the same subjects using the same set of video stimuli and experimental protocol. This database can serve as a valuable research resource for scholars in the field of affective computing. It offers opportunities to investigate emotion recognition and the underlying mechanisms of both wet electrode and dry electrode systems. Additionally, it enables the exploration of the relationship between physiological signal patterns captured by these two types of electrodes. Moreover, the database goes beyond single modality research, as it includes simultaneous collection of peripheral physiological signals such as ECG, EOG, GSR, BVP, RSP and SKT signals. This comprehensive dataset can provide a foundation for conducting research on multimodal emotion recognition, whether using dry electrodes or wet electrodes. By incorporating multiple physiological signals, researchers can further enhance the accuracy and robustness of emotion recognition systems.

Indeed, it must be acknowledged that the achieved performance in dry electrode emotion recognition within this study still leaves ample room for enhancement, which may be attributed to the fact that the current encoder architectures may not be able to fully explore the entire range of EEG features, thereby constraining their capacity to extract the valuable information embedded in wet EEG data effectively. To address this issue, future investigations could concentrate on refining the architecture and configuration of encoders optimized for extracting wet electrode EEG features, which may involve exploring more advanced and complex neural network architectures.


\section{Conclusion}
In this study, we introduce a denoising encoder via contrastive learning alignment network (DECAN) to enhance the performance of dry electrode EEG emotion recognition with the help of wet electrode EEG signals: (1) We propose the DECAN model, a model consisting of two partially shared DNN models and a feature alignment contrastive learning strategy to extract efficient dry electrode EEG emotion features. (2) We construct a new dataset named PaDWEED to better support our research, which contains paired dry and wet electrode EEG datasets from 16 subjects, along with peripheral physiological signals such as ECG, EOG, RSP, GSR, BVP, and SKT. (3) Experimental results on the PaDWEED dataset demonstrate that our proposed model achieves state-of-the-art performance on
the dry electrode EEG emotion recognition task. Additionally, experiments involving feature alignment across subjects and datasets reveal that DECAN can effectively overcome more challenging scenarios.

\bibliographystyle{IEEEtran}
\bibliography{Ref}

\vfill
\end{CJK}
\end{document}